\documentclass[aps,twocolumn,english,superscriptaddress,citeautoscript,showkeys,preprintnumbers,amsmath,amssymb,floatfix,footinbib,prb]{revtex4-2}
\usepackage{xcolor}
\usepackage{soul}
\usepackage{amssymb,amsmath}
\usepackage{hyperref}
\hypersetup{colorlinks=true,linkcolor=red,citecolor=blue}
\usepackage[utf8]{inputenc}
\usepackage[english]{babel}
\usepackage{txfonts}

\usepackage{graphicx} 
\usepackage{dcolumn} 
\usepackage{bm} 
\usepackage{float}
\usepackage{hyperref}

\begin{document}


\title{Superconducting Gap Structure of Filled Skutterudite LaOs$_4$As$_{12}$ Compound through $\mu$SR Investigations}

\author{A. Bhattacharyya}
\email{amitava.bhattacharyya@rkmvu.ac.in} 
\address{Department of Physics, Ramakrishna Mission Vivekananda Educational and Research Institute, Belur Math, Howrah 711202, West Bengal, India}

\author{D. T. Adroja} 
\email{devashibhai.adroja@stfc.ac.uk} 
\affiliation{ISIS Facility, Rutherford Appleton Laboratory, Chilton, Didcot Oxon, OX11 0QX, United Kingdom} 
\affiliation{Highly Correlated Matter Research Group, Physics Department, University of Johannesburg, PO Box 524, Auckland Park 2006, South Africa}

\author{A. D. Hillier}
\affiliation{ISIS Facility, Rutherford Appleton Laboratory, Chilton, Didcot Oxon, OX11 0QX, United Kingdom}

\author{P. K. Biswas} 
\email{Deceased} 
\affiliation{ISIS Facility, Rutherford Appleton Laboratory, Chilton, Didcot Oxon, OX11 0QX, United Kingdom}

\begin{abstract}

Filled skutterudite compounds have gained attention recently as an innovative platforms for studying intriguing low-temperature superconducting properties. Regarding the symmetry of the superconducting gap, contradicting findings from several experiments have been made for LaRu$_{4}$As$_{12}$ and its isoelectronic counterpart, LaOs$_{4}$As$_{12}$. In this vein, we report comprehensive bulk and microscopic results on LaOs$_{4}$As$_{12}$ utilizing specific heat analysis and muon-spin rotation/relaxation ($\mu$SR) measurements. Bulk superconductivity with $T_C$ = 3.2 K was confirmed by heat capacity. The superconducting ground state of the filled-skutterudite LaOs$_{4}$As$_{12}$ compound is found to have two key characteristics: superfluid density exhibits saturation type behavior at low temperature, which points to a fully gapped superconductivity with gap value of $2\Delta/k_BT_C$ = 3.26; additionally, the superconducting state does not show any sign of spontaneous magnetic field, supporting the preservation of time-reversal symmetry. These results open the door for the development of La-based skutterudites as special probes for examining the interplay of single- and multiband superconductivity in classical electron--phonon systems.  

\end{abstract}

\date{\today} 


\maketitle

\section{Introduction}

\noindent \noindent Due to their potential as thermoelectric materials for either refrigeration or power generation applications, many filled skutterudite compounds with RT$_4$X$_{12}$ stoichiometry \linebreak (R = alkali metals, alkaline earth metals, lanthanides, or light actinides; T = Fe, Os, Ru; \linebreak X = P, As, Sb) have lately been the focus of several investigations~\cite{baumbach2010filled,sales1996filled,keppens1998localized}. With two formula units RT$_{4}$X$_{12}$ per unit cell, these compounds form a body-centered cubic structure (space group $Im\bar{3}$, No: 204). The structures consist of rigid covalently bonded cage-forming frameworks T$_4$X$_{12}$ that encapsulate various bonded guest atoms R. This leads to local anharmonic thermal vibrations (rattling modes), which would reduce phononic heat conduction and open the door to their potential as promising thermoelectric materials. Because of the significant hybridization between the $4f$ band manifold and electronic conduction states, as well as the degree of freedom provided by the R-$f$-derived multipole momenta of the cubically symmetric X$_{12}$ cages, those compounds may include a variety of distinct electronic and magnetic ground states. For examples, consider unconventional superconductivity~\cite{bauer2002superconductivity,kotegawa2003evidence,adroja2005probing,zhang2013multiband,kawamura2018filled}, Kondo effect~\cite{takeda2000superconducting,adroja2003spin,adroja2007observation,baumbach2008filled,shankar2019ceos4as12}, heavy fermios~\cite{sanada2005exotic}, non-Fermi liquid behavior~\cite{takeda2000superconducting}, etc.\\

 \noindent The majority of the Pr- and Ce-based filled skutterudite compounds are hybridized gap semiconductors or show magnetic transitions, however PrOs$_{4}$Sb$_{12}$~\cite{bauer2002superconductivity,kotegawa2003evidence}, PrRu$_{4}$Sb$_{12}$~\cite{adroja2005probing} and PrRu$_{4}$As$_{12}$~\cite{shirotani1997superconductivity} show superconducting transitions at 1.8 K, 0.97 K and 2.4 K, respectively. PrOs$_4$Sb$_{12}$ is highly intriguing for a variety of reasons~\cite{maple2006unconventional}, including: (i) it is the first known example of a heavy-fermion superconductor containing Pr; (ii) it shows unconventional strong-coupling superconductivity that breaks time-reversal symmetry; and (iii) instead of magnetic fluctuations, electric quadrupole fluctuations may be involved in the superconducting pairing process. The unique band structure of these compounds and the hybridization effects between localized $f$ electrons and conduction electrons appear to play a crucial role, in addition to the fact that the origin of the majority of those unconventional phenomenologies is unknown. It was recently revealed that the Fermi level of La compounds is placed at a prominent peak arising from the T-$d$ band manifold, which might contribute to electronic instability~\cite{baumbach2010filled,nordstrom1996electronic}. Several La-based compounds LaT$_4$X$_{12}$ are especially remarkable within the filled skutterudite class due to their remarkable superconducting properties. For examples, LaFe$_{4}$P$_{12}$ ($T_{C}$ = 4.1 K)~\cite{meisner1981superconductivity}, LaOs$_{4}$P$_{12}$ ($T_{C}$ = 1.8 K)~\cite{meisner1981superconductivity,shirotani1996electrical}, and LaRu$_{4}$Sb$_{12}$ ($T_{C}$ = 3.6 K)~\cite{uchiumi1999superconductivity,takeda2000superconducting}, with a special attention to the LaRu$_{4}$As$_{12}$ ($T_{C}$ = 10.3 K, $H_{c2}$ = 10.2 T)- with the highest superconducting transition temperature.~\cite{shirotani1996electrical,shirotani1997superconductivity,shirotani2000electrical}.\\ 
 
\noindent The ratio of the heat capacity jump $\Delta C$ to $\gamma$T$_C$ is $\Delta$C/($\gamma$T$_C$)=1.75 for LaRu$_{4}$As$_{12}$ comparison to the BCS value of 1.43~\cite{shirotani1997superconductivity}. While the majority of La-based filled skutterudites are completely gapped superconductors, past research has shown numerous unique aspects of LaRu$_{4}$As$_{12}$, such as a positive curvature of $H_{c2}$, nonexponential behavior of the electronic heat capacity, and square root field dependency of the Sommerfeld coefficient ($\gamma$)~\cite{matsuhira2009systematic}. We recently reported unambiguous evidence of multiband $s+s$-wave superconductivity in LaRu$_{4}$As$_{12}$ using muon-spin rotation measurements, with $2\Delta_1/k_BT_C $ = 3.73 for the larger gap and $2\Delta_ 2/k_BT_C $ = 0.144 for the smaller gap~\cite{bhattacharyya2022multigap}. Furthermore, inelastic X-ray scattering experiments indicated essentially temperature-independent phonon modes between 300 K and 20 K, with the exception of 2 K, where a weak softening of the specific phonon modes is detected~\cite{bhattacharyya2022multigap}. All of these results demonstrate the relevance of the electron--phonon interaction in the superconductivity of LaRu$_{4}$As$_{12}$, and they accord well with the DFT-based phonon simulations~\cite{bhattacharyya2019investigation}.\\

\noindent Another isostructural La-based filled skutterudite compound, LaOs$_{4}$As$_{12}$, has been reported by Shirotani et\,al. to exhibit superconductivity with $T_C$. = 3.2 K~\cite{shirotani2000electrical}. LaOs$_{4}$As$_{12}$ has also shown some signs of multiband superconductivity, such as the upward curving of the upper critical field around the transition temperature and unusual behavior in the electronic specific heat data~\cite{juraszek2016specific}.~A single-gap, s-wave superconducting ground state, however, is suggested by a recent study of the temperature dependency of lower critical field~\cite{juraszek2020symmetry}. Another study found that the high-amplitude lanthanum phonons dominate the vibrational eigenmodes at low energies based on the phonon dispersion relation determined from inelastic neutron scattering experiments~\cite{marek2013vibrational}.\\

\noindent We have thus performed systematic muon-spin rotation and relaxation ($\mu$SR) measurements to examine the superconducting pairing process in the LaOs$_{4}$As$_{12}$ compound. Contrary to prior experimental work asserting two-band superconductivity~\cite{juraszek2016specific}, we demonstrate that the low-temperature behavior of the superfluid density points to a fully gapped superconducting Fermi surface. Furthermore, the preservation of time-reversal symmetry is confirmed by the lack of spontaneous magnetic fields in the superconducting state, ruling out unusual pairing processes. The transition from two-band to single-band superconductivity in LaRu$_{4}$As$_{12}$ to LaOs$_{4}$As$_{12}$ is caused by differences in interband coupling strength in the Fermi surface, as evidenced by the different degrees of hybridization and electronic properties observed in the Fermi surfaces of both compounds~\cite{Ferreiratheory}. These results underline the significance of LaRu$_{4}$As$_{12}$ and LaOs$_{4}$As$_{12}$ compounds as an important platform for investigating filled skutterudites for the competition between single-band and multiband superconductivity in electron--phonon driven systems.\\ 

\begin{figure}[t]
 
\includegraphics[width= \linewidth]{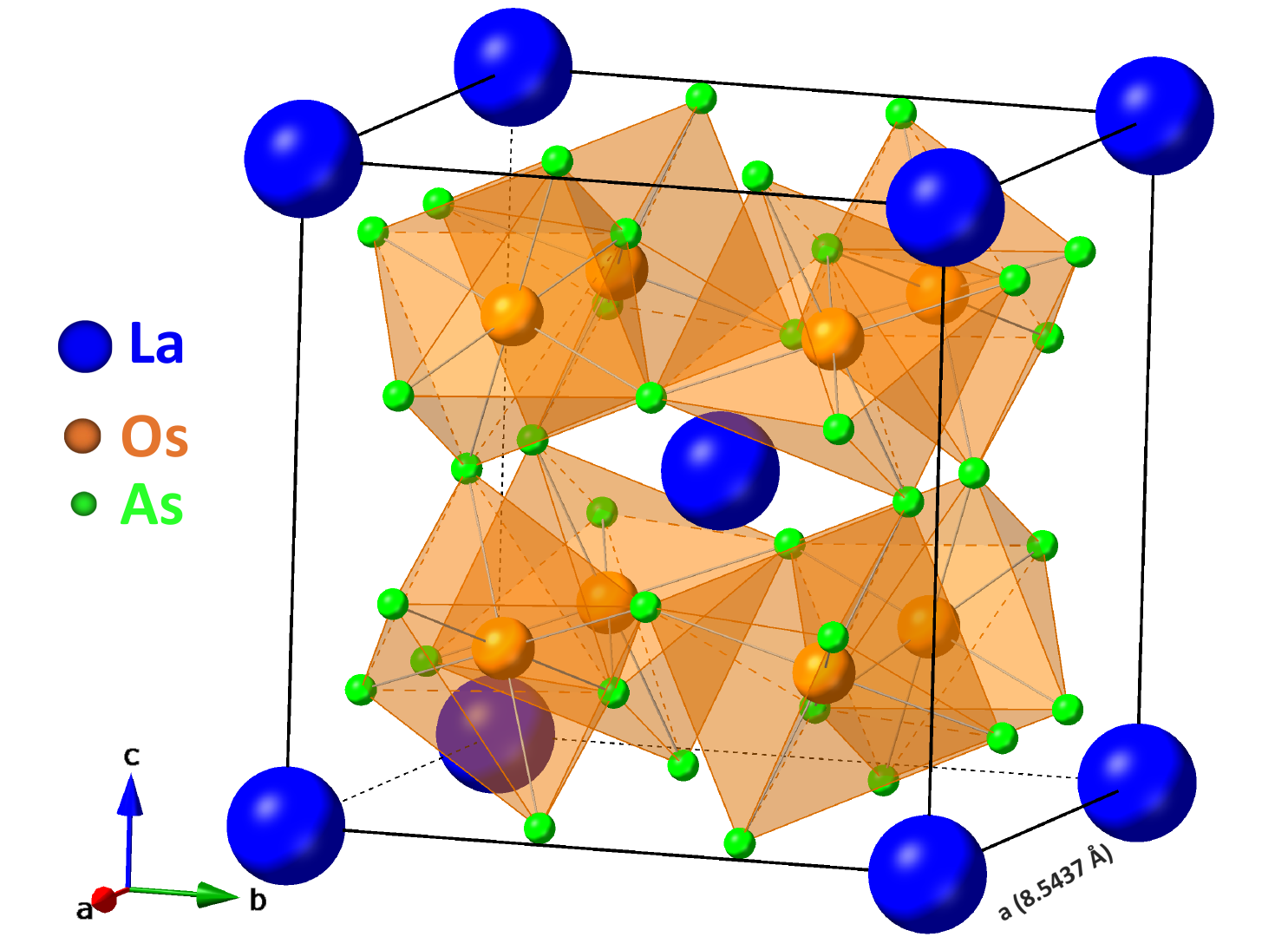}
\caption{A unit cell of the body-centered cubic LaOs$_4$As$_{12}$ structure with the space group $Im\bar{3}$ that crystallizes within a CoAs$_3$-type skutterudite structure packed with La atoms. Green: As, Orange: Os, and Blue: La.} 
\label{crystal structure}
\end{figure}

\begin{figure}[h]
 
\includegraphics[height=1.35\linewidth, width=0.9\linewidth]{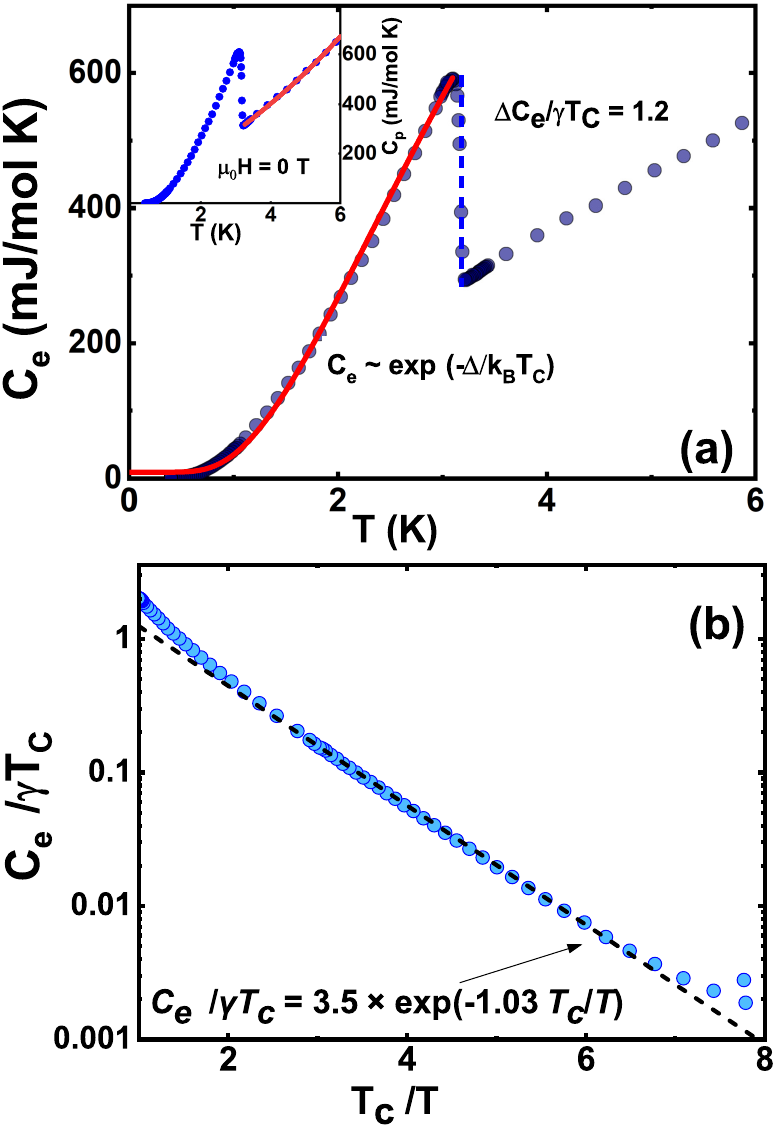}
\caption{\hl{(\textbf{a})} Low-temperature specific heat of the filled skutterudite compound LaOs$_4$As$_{12}$, expressed as $C_e$ vs $T$ in a zero magnetic field. The specific heat data is shown in the inset. (\textbf{b}) The normalized electronic specific heat ($C_{e}/\gamma T_{C}$) versus inverse reduced temperature ($T_C/T$). The heat capacity data are from~\cite{juraszek2016specific}.} 
\label{resistivity and heat capacity}
\end{figure}

\section{Experimental Details}

\noindent  The high-temperature molten-metal-flux technique, described in~\cite{henkie2008crystal}, was used to grow single crystals of LaOs$_4$As$_{12}$. In a quartz ampule, elements with purities higher than 99.9\% and a molar ratio of La:Os:Cd:As $\rightarrow$ 1:4:12:48 were combined. The details on the single crystal growth can be found in \cite{henkie2008crystal}. The relaxation approach was used to measure the heat capacity in a Quantum Design physical properties measurement (PPMS) system. Temperatures as low as 0.38 K were attained utilizing a He-3 attachment to the PPMS~\cite{juraszek2016specific}.\\

\par

The $\mu$SR measurements were carried out using small size unaligned single crystals (0.1 mm \hl{$\times$}  0.1 mm \hl{$\times$} 0.1 mm, total mass 1 $\rm{g}$), which gave powder average muon signal, of LaOs$_4$As$_{12}$. The MuSR spectrometer at the Rutherford Appleton Laboratory, ISIS Neutron and Muon Source in the UK was used to perform the $\mu$SR measurements~\cite{lee1999muon}. In a $\mu$SR experiment, the sample is injected with 100\% spin-polarized muons. Each implanted muon thermalizes, at which point it decays (lifetime $\tau_{\mu}$ = 2.2 $\mu$s) into a positron (and two neutrinos) which is preferentially released in the direction of the muon spin at the moment of decay. Utilizing detectors carefully placed around the sample, the decay positrons are detected and time-stamped. It is possible to calculate the asymmetry in the positron emission as a function of time, $A(t)$, using the collected histograms from the forward (F) and backward (B) detectors, $A(t)=\frac{N_{\mathrm{F}}(t)-\alpha N_{\mathrm{B}}(t)}{N_{\mathrm{F}}(t)+\alpha N_{\mathrm{B}}(t)}$, where $\alpha$ is a calibration factor for the instrument and $N_{\mathrm{F}}(t)$ and $N_{\mathrm{B}}(t)$ are the number of positrons counted in the forward and backward detectors, respectively. Detectors are placed longitudinally during ZF-$\mu$SR, and a correction coil is used to cancel out any stray magnetic fields up to 10$^{-4}$~mT. To investigate the time reversal symmetry ZF-$\mu$SR measurements were carried out~\cite{sonier2000musr}. In the vortex state, TF-$\mu$SR measurements were performed with applied fields of 20, 30, 40, 50, and 60~mT, which is greater than the lower critical field $H_\mathrm{c1}$ ($\sim$5~mT) and lower than the upper critical field $H_\mathrm{c2}$ ($\sim$1\,T)~\cite{shirotani2000electrical}. The sample was covered using a thin silver foil after being mounted onto a high purity (99.995\%) silver sample holder using diluted GE-varnish. The sample was cool down to 300 mK using a dilution refrigerator. To generate the vertex lattice by trapping the applied TF, we applied field above $T_{\mathrm{C}}$ and then sample was cooled in the field to the base temperature of 300 mK. We used WiMDA~\cite{pratt2000wimda} software to analyze the $\mu$SR data.\\

\section{Results and discussion}

\subsection{Crystal Structure \& Physical Properties}

\noindent LaOs$_4$As$_{12}$ crystallizes in a CoAs$_3$-type skutterudite structure packed with La atoms and has a body-centered cubic structure with the space group $Im\bar{3}$ (No. 204) as shown in Figure~\ref{crystal structure}. The large icosahedron cage made of As atoms is located around the electropositive La sites, which lack four-fold rotational symmetry. Between the cages, a transition metal ion called Os forms a cubic sublattice. The low temperature specific heat measurements $C_{P}$ as a function of temperature at zero magnetic field are shown in the inset of Figure~\ref{resistivity and heat capacity}a. Using the equations $C_{P} = \gamma T + \beta T^{3}$, the normal state heat capacity is fitted. We calculated the lattice contribution to the specific heat $\beta$ = 0.613 mJ/mol K$^{4}$ and the electronic parameter (Sommerfeld's coefficient) $\gamma$ = 90.47 mJ/mol K$^{2}$ from this. The Debye temperature is determined using the Debye model as $\Theta_{D} = \left(\frac{12\pi^{4}nR}{5\beta}\right)^{1/3}$, where $R$ is the universal gas constant, which is 8.31\hl{4} 
 J/mol-K, and $n$ denotes the number of atoms in the compound (n = 17). The value of $\Theta_{D}$ is thus calculated to be approximately 377 K, which agrees with the previous measurement~\cite{juraszek2016specific,matsuhira2009systematic}. Figure~\ref{resistivity and heat capacity}a displays the low-$T$ electronic specific heat $C_e$ that was produced after the phonon contribution was taken into account. The heat capacity jump at $T_{C}$ ($\Delta C_{e}/\gamma T_{C}$) is calculated to be 1.2, which is less than 1.43 the value expected for a weak-coupling BCS superconductivity. The fit to the exponential temperature dependency of $C_{e}(T)$ yields $\Delta(0) = 0.40$ meV, which is close to the 0.45 meV value obtained from the TF-$\mu$SR data analysis (discussed in section-B). Thus, the value of $2\Delta(0)/k_BT_C$ = 2.9, which is less than the 3.53 anticipated for weak-coupling BCS superconductors. However, the linear fitting shown in Figure~\ref{resistivity and heat capacity}b shows that this material exhibits BCS behavior with a single isotropic gap.\\

\begin{figure*}[t]
 
\includegraphics[height=0.5\linewidth, width=0.5\linewidth]{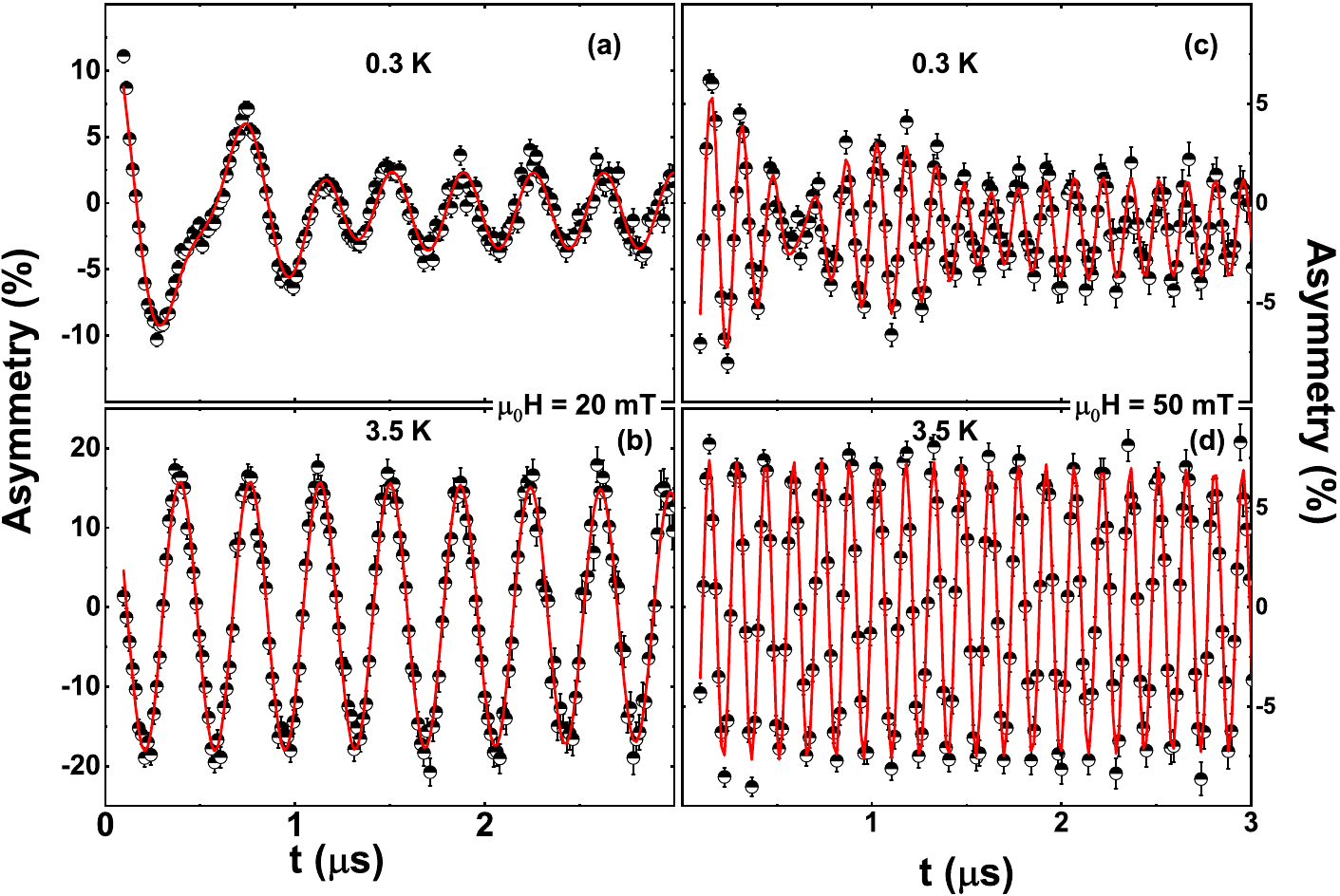}
\includegraphics[height=0.5\linewidth,width=0.45\linewidth]{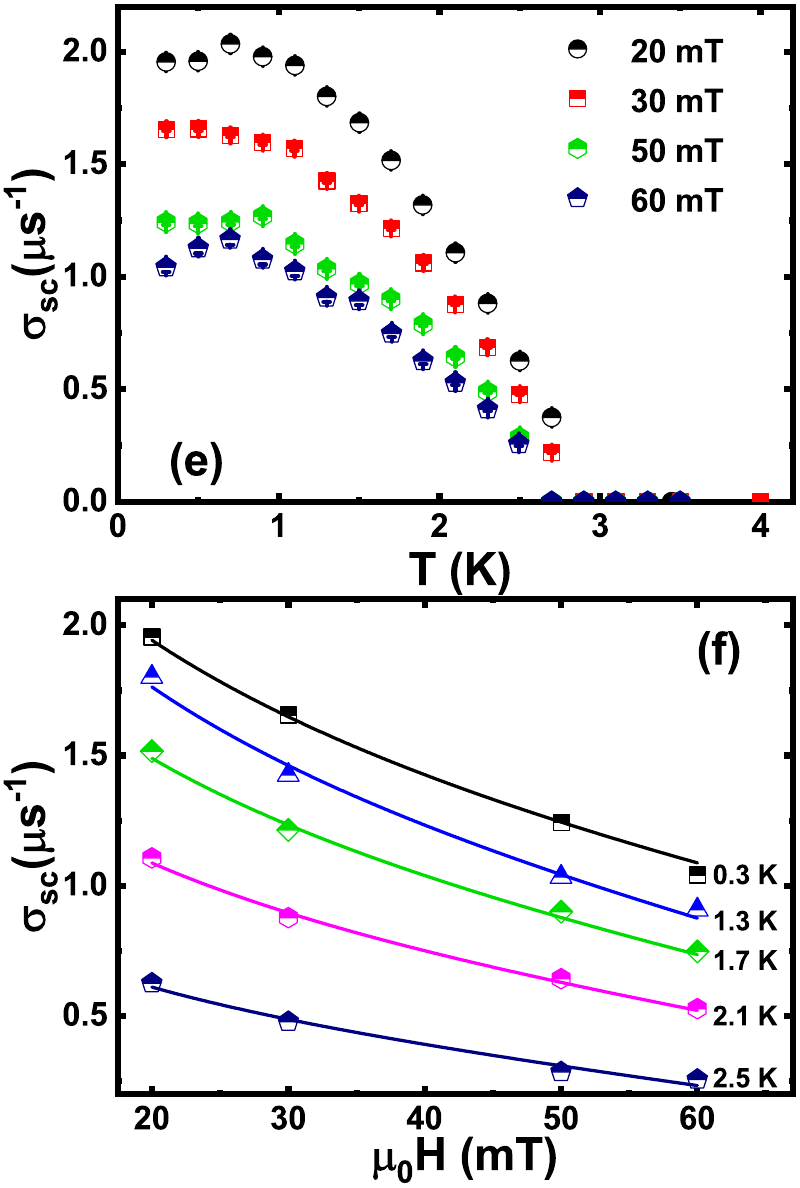}
\caption {{\bf Left panel:} \hl{Asymmetry} spectra of the TF-$\mu$SR in the low time region obtained in 20 mT and 50 mT applied magnetic fields at (\textbf{a}--\textbf{c}) $T $ = 0.3 K (i.e., below $T_\mathrm{C}$) and (\textbf{b}--\textbf{d}) $T $ = 3.5 K (i.e., above $T_\mathrm{C}$). {\bf Right panel:} (\textbf{e}) The superconducting depolarization rate $\sigma_{sc}$ as a function of temperature in the presence of an applied field of 20 $\leq\mu_{0}$ H $\leq$ 60 mT. (\textbf{f}) The magnetic field dependence of the muon spin depolarization rate is shown for a range of different temperatures. The solid lines are the results of fitting the data using Brandt’s equation as discussed in Equation (2).}
\label{TF Asymmetry and depolarization rate} 
\end{figure*}

\subsection{Superconducting Gap Structure: TF-$\mu$SR}

\noindent The pairing mechanism and superconducting gap structure of the LaOs$_4$As$_{12}$ were investigated by TF-$\mu$SR experiments down to 0.3 K. The TF-$\mu$SR asymmetry time spectra in the presence of 20 mT and 50 mT applied magnetic fields at above and below $T_\mathrm{C}$ are shown in Figures~\ref{TF Asymmetry and depolarization rate}a--d. Because of the extra inhomogeneous field distribution of the vortex lattice generated inside the superconducting mixed state of LaOs$_4$As$_{12}$, the spectrum in Figure~\ref{TF Asymmetry and depolarization rate}a,c in the superconducting state at 0.3 K demonstrate a greater relaxation. Using the Gaussian damped decay function, the asymmetry spectra were fitted~\cite{Bhattacharyyarev, BhattacharyyaThCoC2,CeIr3} using the following equation, 
\begin{equation}
\begin{split}
A_\mathrm{TF}(t) = A_\mathrm{sc}\exp\left(-\frac{\sigma_{TF}^{2}t^{2}}{2}\right)\cos(\gamma_{\mu}B_{sc}t+\phi) +\\A_\mathrm{bg}\cos(\gamma_{\mu}B_{bg}t+\phi).
\end{split}
\end{equation}

\begin{figure*}[t]
\includegraphics[width=0.45\linewidth, height=0.4\linewidth]{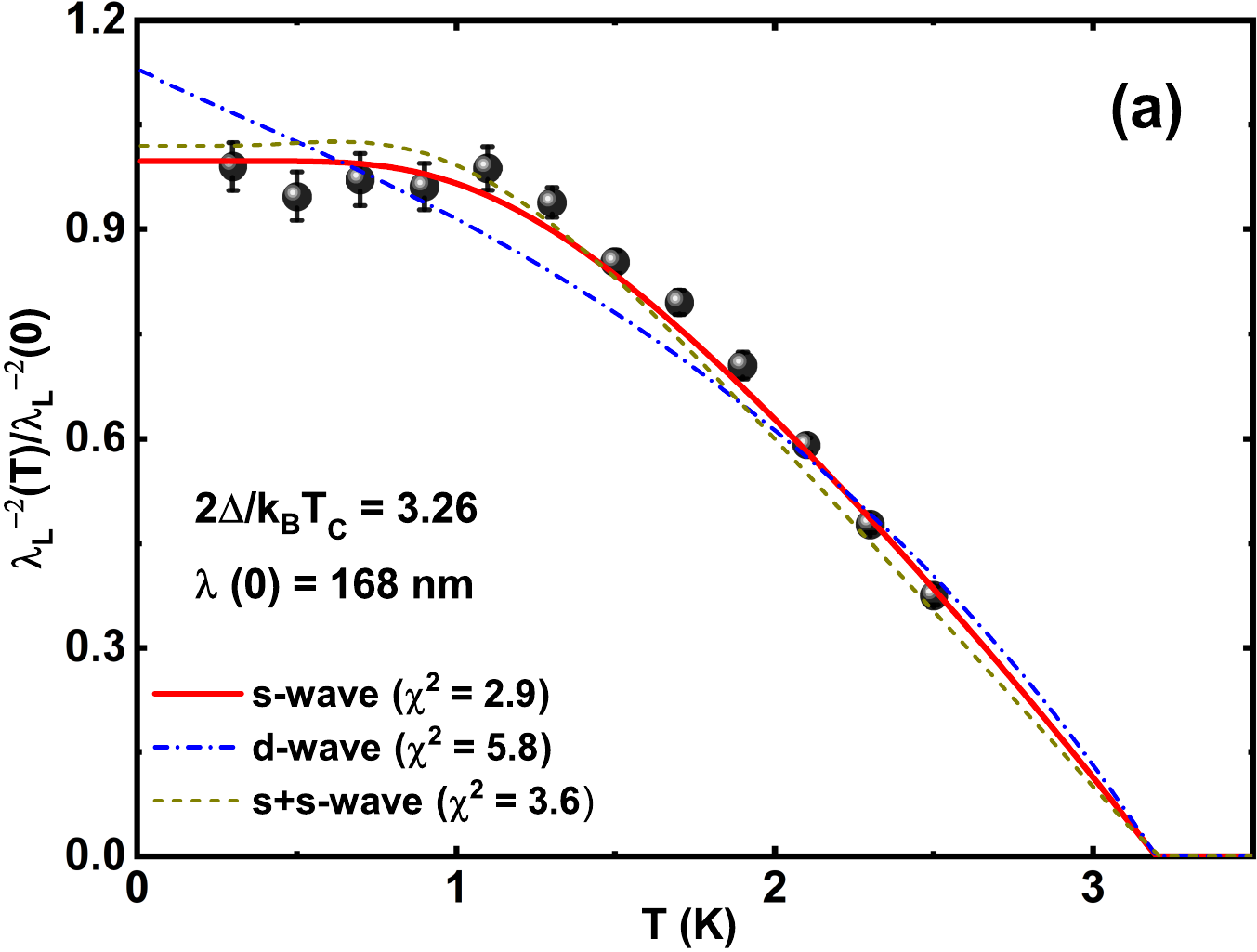}
\includegraphics[width=0.45\linewidth, height=0.4\linewidth]{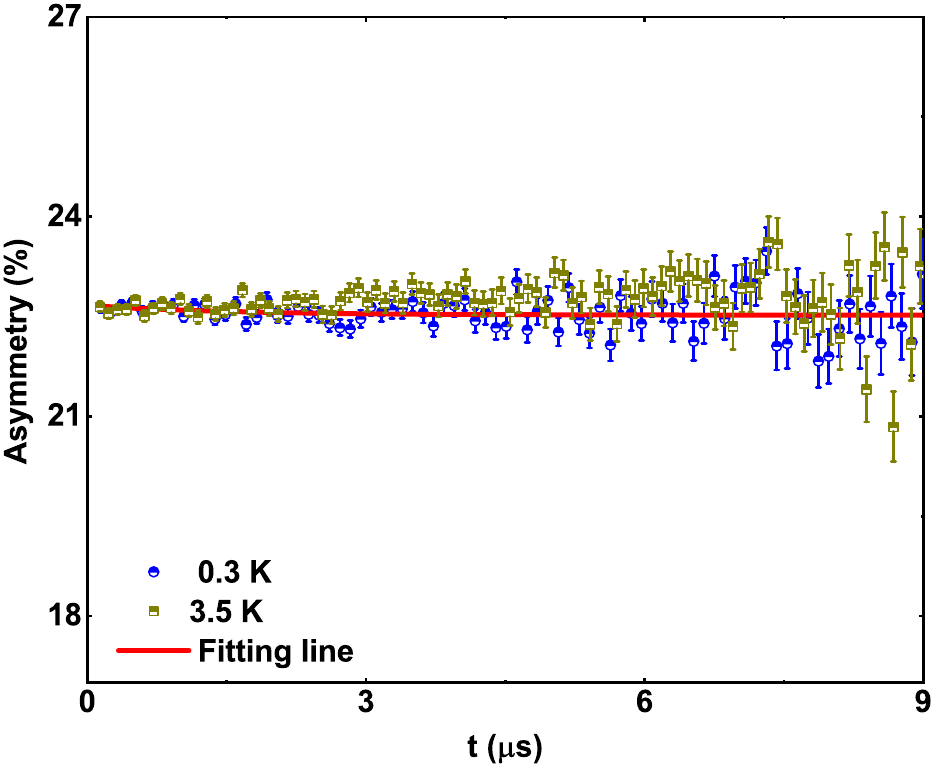}
\caption{{\bf Left panel :} (\textbf{a}) \hl{The} inverse magnetic penetration depth squared as a function of temperature is shown here. The lines show the fits using $s$-wave (red), $s+s$-wave (light green) and $d$-wave (blue) gap functions. {\bf Right panel :} (\textbf{b}) Shows the ZF-$\mu$SR spectra for LaOs$_4$As$_{12}$ at 0.3 K (blue) and 4 K (light green). The solid line fits to the experimental data points, as stated in the text.}
\label{inverse magnetic penetration depth and ZF MuSR} 
\end{figure*}

The gyromagnetic muon ratio is $\gamma_\mu/2\pi$ = 135.53 MHz/T, and the initial asymmetries of muons stopping on the sample and on the silver holder are $A_{sc}$ and $A_{bg}$, respectively (constant across the entire temperature range). The local fields B$_{sc}$ and B$_{bg}$ represent muons stopping on the sample and on the sample holder, respectively, whereas $\phi$ represents initial phase value and $\sigma_{TF}$ represents the Gaussian depolarization rate. We calculated the values of $A_{sc}$ = 76\% and $A_{bg}$ = 24\% of the total asymmetries by fitting 0.3 K data. When additional temperature data were analyzed, $A_{bg}$ was kept constant and $A_{sc}$ was found nearly temperature independent. The emergence of bulk superconductivity is indicated by an increase in the $\sigma_{TF}$ rate as the system approaches the superconducting state. With the use of the following formula, the superconducting contribution to the relaxation $\sigma_{sc}$ was determined, $\sigma_\mathrm{sc} = \sqrt{\sigma_\mathrm{TF}^{2}-\sigma_\mathrm{nm}^2}$, where the nuclear magnetic dipolar contribution, is denoted by the symbol $\sigma_{nm}$, which is derived from high-temperature fits and is temperature independent. Figure~\ref{TF Asymmetry and depolarization rate}e depicts the temperature dependence of $\sigma_{sc}$ in several applied TF fields. Due to low $H_\mathrm{c2}$ value, as seen in Figure~\ref{TF Asymmetry and depolarization rate}f, $\sigma_{\mathrm{sc}}$ depends on the applied field. Brandt demonstrated that the London penetration depth $\lambda_{L}(T)$ is linked to $\sigma_\mathrm{sc}$ for a superconductor with $H_\mathrm{ext}/H_\mathrm{c2}$ $\leq$ 0.25~\cite{brandt1988magnetic,brandt2003properties}.
\begin{equation}
\begin{split}
\sigma_\mathrm{sc}[\mu s^{-1}] = 4.83 \times 10^{4}(1-H_\mathrm{ext}/H_\mathrm{c2}) \\ 
\times \{1+1.21\left[1-\sqrt{{(H_\mathrm{ext}/H_\mathrm{c2}})}\right]^{3}\}\lambda_{L}^{-2}[nm].
\end{split}
\label{eqn2}
\end{equation}

 This relationship has been used to compute the temperature dependency of $\lambda_{L}(T)$. As demonstrated in Figure~\ref{TF Asymmetry and depolarization rate}f, isothermal cuts perpendicular to the temperature axis of $\sigma_{sc}$ data sets were utilized to estimate the $H$-dependence of the depolarization rate $\sigma_{sc}(H)$. The normalized $\lambda_{L}^{-2}(T)/\lambda_{L}^{-2}(0)$ temperature variation, which is directly proportional to superfluid density, is shown in Figure \ref{inverse magnetic penetration depth and ZF MuSR}a. The data were fitted using the following equation~\cite{Prozorov,AdrojaK2Cr3As3}: 
 \vspace{-3pt}
\begin{eqnarray}
\frac{\sigma_{sc}(T)}{\sigma_{sc}(0)} &=& \frac{\lambda_{L}^{-2}(T)}{\lambda_{L}^{-2}(0)}\\
 &=& 1 + \frac{1}{\pi}\int_{0}^{2\pi}\int_{\Delta(T)}^{\infty}\left(\frac{\delta f}{\delta E}\right) \times \frac{EdEd\phi}{\sqrt{E^{2}-\Delta(T,\phi})^2}, \nonumber
\end{eqnarray}

\noindent where $f = [1+\exp(\frac{E}{k_{B}T})]^{-1}$ is the Fermi function. We take $\Delta_{k}(T,\phi) = \Delta(T)\mathrm{g}_{k}(\phi)$, where we assume a temperature dependence that is universal $\Delta(T) = \Delta_0 \tanh[1.82\big\{1.018(T_\mathrm{C}/T-1)\big\}^{0.51}]$. The magnitude of the gap at 0 K is $\Delta_0$, and the function $\mathrm{g}_{k}$ denotes the gap's angular dependence, which is equal to 1 for one isotropic energy gap $s$, 1 for two isotropic $s+s$ wave energy gap and $\cos(2\phi)$ for d-wave gap, where $\phi$ is the azimuthal angle along the Fermi surface.\\

\par

\noindent Figure~\ref{inverse magnetic penetration depth and ZF MuSR}a illustrates our comparison of three distinct gap models: employing a single isotropic $s$-gap wave, a multigap $s+s$-wave gap, and a nodal $d$-wave gap. As seen in the figure, the superfluid density saturates at low temperatures, which is a characteristic of the $s$-wave model with a single gap. An isotropic single-band $s-$wave model with a gap value of 0.45 meV provides the best representation of the data, with a gap to $T_\mathrm{C}$ ratio $2\Delta(0)/k_\mathrm{B}T_\mathrm{C}$ = 3.26, which is less than the BCS weak-coupling limit (=3.53). On the other hand, the substantial rise in the $\chi^2$ value puts the $d$-wave model and $s+s$-wave (multigap) model inappropriate for this system. A two-gap $s$+$s$-wave model of multiband superconductivity has been shown to be compatible with the temperature dependence of magnetic penetration depth of LaRu$_4$As$_{12}$. The higher gap to $T_C$ ratio computed in the $s + s$-wave scenario, 2$\Delta_1(0)/k_{\mathrm{B}}T_C$ = 3.73, is fairly comparable to the value of 3.53 for BCS superconductor in case of LaRu$_4$As$_{12}$~\cite{bhattacharyya2022multigap}. For LaRu$_4$As$_{12}$, 2 K specific phonon modes exhibit modest softening when compared to 20 K, demonstrating that the electron--phonon interactions causing the superconductivity have an audible impact on the vibrational eigenstates~\cite{bhattacharyya2022multigap}. Using McMillan's relation, it is also possible to determine the electron--phonon coupling constant ($\lambda_\mathrm{e-ph}$)~\cite{McMillan}:
\begin{equation}
\lambda_\mathrm{e-ph} = \frac{1.04+\mu^{*}\ln(\Theta_\mathrm{D}/1.45T_\mathrm{C})}{(1-0.62\mu^{*})\ln(\Theta_\mathrm{D}/1.45T_\mathrm{C})-1.04}.
\end{equation}
where $\mu^*$ is the repulsive screened Coulomb parameter usually assigned as $\mu^*$ = 0.13. The calculated value of the $\lambda_\mathrm{e-ph}$ is 0.534. The London model is described as $\lambda_\mathrm{L}^2=m^{*}c^{2}/4\pi n_\mathrm{s} e^2$. It connects the effective mass enhancement m$^*$ [$ =(1+\lambda_{e-ph})*m_e]$, superconducting carrier density $n_{\mathrm{s}}$ [$ =m^{*}c^{2}/4\pi e^2\lambda_{L}(0)^2$], and London penetration depth. By employing the $s$-wave model, we determined the London penetration depth of $\lambda_L(0)$ = 168 nm. The effective mass enhancement is calculated to be $m^{*} = 1.53~m_\mathrm{e}$, and the superconducting carrier density is predicted to be $n_\mathrm{s} = 1.53 \times 10^{27}$ carriers m$^{-3}$. References~\cite{Chia,Amato,bhattacharyya2019investigation} include a description of the computations in detail. The calculated values of \mbox{$\lambda_L(0)$ = 240 nm},\linebreak $n_\mathrm{s} = 8.6 \times 10^{27}$ carriers m$^{-3}$ and $m^{*} = 1.749~m_\mathrm{e}$ for LaRu$_4$As$_{12}$~\cite{bhattacharyya2022multigap}. The fitted parameters for LaOs$_{4}$As$_{12}$ and LaRu$_{4}$As$_{12}$ (for comparison) are shown in Table \ref{Table}.~To explain the observed nature of the superconducting gap structures, it is important to comprehend the electronic structures of these compounds, which have been carried~\cite{Ferreiratheory} and the results suggest that the single-band order parameter in
LaOs$_{4}$As$_{12}$ seems to be associated with the hybridized As-p and Os-d electronic character
of the Fermi surface. On the other hand, the lack of hybridization for the disjointed Fermi surface of LaRu$_{4}$As$_{12}$, may explain its multiband superconducting nature.\\

\begin{table}[htbp]
  \caption{Fitted parameters for LaOs$_{4}$As$_{12}$ and LaRu$_{4}$As$_{12}$.}
  \label{Table}
  \centering
  \renewcommand{\arraystretch}{1.2} 
  \begin{tabular}{cccccc}
    \hline\hline
    \textbf{Model} & $\Delta_{i}(0)$ (meV) & $2\Delta_{i}(0)/k_{B}T_{C}$ & $T_{C}$ (K) & $\lambda_{L}(0)$ (nm) \\
    \hline
    LaOs$_{4}$As$_{12}$ (s-wave) & 0.45 & 3.26 & 3.2 & 168 \\
    LaRu$_{4}$As$_{12}$ (s + s-wave) & 1.656, 0.064 & 3.73, 0.144 & 10.3 & 240 \\
    \hline
  \end{tabular}
\end{table}

\subsection{Preserved Time Reversal Symmetry: ZF-$\mu$SR}

\noindent In order to determine if there is a spontaneous magnetic field present in the superconducting ground state, we conducted the ZF-$\mu$SR experiment. Figure~\ref{inverse magnetic penetration depth and ZF MuSR}b shows the time evolution of the asymmetry spectra for $T$ = 0.3\,K $< T_\mathrm{C}$ and $T$ = 3.5\,K $> T_\mathrm{C}$. The ZF-$\mu$SR spectra recorded in the normal and superconducting states show the same relaxations that can be found in overlapping ZF-$\mu$SR spectra, indicating that the superconducting state does not shows any spontaneous magnetic field or spin fluctuations. This result suggests that the time-reversal symmetry is preserved in LaOs$_{4}$As$_{12}$ superconducting state. The strong resemblance of the ZF-$\mu$SR spectra (above and below T$_C$) suggests that the time-reversal symmetry is also retained in the superconducting state of LaRu$_{4}$As$_{12}$. In order to fit the ZF data, a Lorentzian function was used~\cite{ZrIrSi}, 
\begin{equation}
G_\mathrm{ZF}(t) = A_\mathrm{sc}(t)\exp{(-\lambda_{ZF} t)}+A_\mathrm{bg},
\end{equation} 
where $\lambda_{ZF}$ is the electronic relaxation rate, $A_\mathrm{sc}$ stands for the sample asymmetry, $A_\mathrm{bg}$ for the constant nondecaying background signal. The red line in Figure~\ref{inverse magnetic penetration depth and ZF MuSR}b indicates the fits to the ZF-$\mu$SR data. The ZF-$\mu$SR asymmetry data fitting parameters are $\lambda_{ZF}$ = 0.754(4) $\mu \mathrm{s}^{-1}$ at 0.3 K and $\lambda_{ZF}$ = 0.744(5) $\mu \mathrm{s}^{-1}$ at 3.5\,K. No conclusive evidence of TRS breaking can be found since the relaxation rate change is within the error bar.\\

\section{Summary}

\noindent We employed TF-$\mu$SR to determine the gap symmetry of the superconducting state of LaOs$_4$As$_{12}$. An isotropic BCS-type $s$-wave gap model explains the temperature dependence of the superfluid density. The gap to $T_{\mathrm{C}}$ ratio, which was determined from the $s$-wave gap fit to the superfluid density, is 3.26; nonetheless, this is smaller than 3.53 expected for conventional BCS systems.~The ZF-$\mu$SR spectra at 0.3 K and 3.5 K are strikingly similar, indicating that the time-reversal symmetry is intact. These results open up the possibility of using the compounds LaRu$_4$As$_{12}$ and LaOs$_4$As$_{12}$ as special research platforms for investigating filled skutterudites for the interplay between single- and multiband superconducting order parameters in conventional systems.\\   

\subsection*{Acknowledgements}

\noindent We thank T. Cichorek and J. Juraszek for providing LaOs$_4$As$_{12}$ sample and the ascii heat capacity data. We would like to thank T. Cichorek, P. P. Ferreira, R. Lucrezi, J. Juraszek, C. Heil and L. T. F. Eleno for interesting discussions. AB expresses gratitude to the Science and Engineering Research Board for the CRG Research Grant (CRG/2020/000698 \& CRG/2022/008528) and CRS Project Proposal at UGC-DAE CSR (CRS/2021-22/03/549). DTA appreciates the support provided by the Royal Society of London for the Newton Advanced Fellowship between the UK and China, the International Exchange between the UK and Japan, and EPSRC-UK (Grant number EP/W00562X/1). We thanks the ISIS Facility for the beam time, RB1520431~\cite{MuSRdata}.\\

\bibliography{laos4as12}
\bibliographystyle{apsrev4-1}

\end{document}